\documentclass[aps, reprint,prl,superscriptaddress,nofootinbib]{revtex4-1}
\usepackage{dcolumn}
\usepackage[utf8]{inputenc}
\usepackage{amsmath}
\usepackage{amsfonts}
\usepackage{amssymb}
\usepackage{graphicx}
\usepackage{hyperref}
\usepackage{xcolor}

\newcommand{\ket}[1]{\ensuremath{|{#1}\rangle}}
\newcommand{\bra}[1]{\ensuremath{\langle{#1}|}}

\usepackage{tikz}
\usetikzlibrary{quantikz}
\begin{document}

\title{Vindication of entanglement-based witnesses of non-classicality in hybrid systems}

\author{Emanuele Marconato}
\affiliation{Dipartimento di Fisica, Università di Torino, Via Pietro Giuria 1, 10125 Torino, Italy}

\author{Chiara Marletto}
\affiliation{Clarendon Laboratory, University of Oxford, Parks Road, Oxford OX1 3PU, United Kingdom}
\affiliation{Centre for Quantum Technologies, National University of Singapore, 3 Science Drive 2, Singapore 117543, Singapore}

\date{\today}%

\begin{abstract}
Recently, Vedral and one of us proposed an entanglement-based witness of non-classicality in systems that need not obey quantum theory, based on constructor-information-theoretic ideas \cite{MVPRD, MAVE}, which offers a robust foundation for recently proposed table-top tests of non-classicality in gravity, \cite{SOUG, MAVE}. The witness asserts that if a mediator can entangle locally two quantum systems, then it has to be non-classical. Hall and Reginatto \cite{HAL, HARE} claimed that there are classical systems that can entangle two quantum systems, thus violating our proposed witness. Here we refute that claim, explaining that the counterexample proposed by Hall and Reginatto in fact validates the witness, vindicating the witness of non-classicality in its full generality. 

\end{abstract}

\maketitle

%\pacs{}

\section{Witnesses of non-classicality}

Quantum theory is one of the best existing explanations of the physical world and it has so far proven unchallenged by experimental evidence. There are however several arguments \cite{Collapse} supporting the idea that quantum theory ceases to apply at a certain scale, for instance due to the collapse of macroscopic quantum superpositions.  It is often suggested that gravity may be responsible for this phenomenon. One of the reasons for this claim is that there is currently no unique viable proposal for a theory of quantum gravity; in addition, the predictions of existing models appear to be beyond current experimental capabilities, \cite{DYS}. The best explanation of gravity is, still, general relativity - which is a classical theory, in the sense that it does not admit quantum superpositions and related effects. 

Recently, an experimental scheme has been proposed in order to refute classical theories of gravity in the presence of quantum masses,\cite{MAVE, SOUG}. This scheme is based on probing gravity with two masses, each in a superposition. If gravity can entangle them, then gravity must be non-classical. As perceptively noted by Hall and Reginatto, \cite{HARE}, it is improper to talk about such tests as proving quantum effects in gravity; what they actually do is to show that, if gravitational entanglement is observed between the masses, gravity must be non-classical - which is a lesser property than having the full set of quantum features.  By {\sl non-classical} we mean that gravity must have at least two variables that are necessary to describe its features, and yet cannot be measured to arbitrarily high accuracy simultaneously (i.e. by the same measuring system). This notion generalises what in quantum theory is called {\sl complementarity} and it can be defined in fully general, information-theoretic terms, \cite{MVPRD}. 

There are two notable features in this proposed witness. One is that it probes a regime where quantum-gravity effects are relevant and measurable, but general relativity's effects are not, because gravity can be described in the Newtonian limit.  The other is that the witness is based on a general theorem that can be proven with minimal assumptions about the dynamics that gravity obeys. This, in short, is the theorem: if a system  ${\bf M}$ (e.g. gravity) can entangle two quantum systems ${\bf Q_A}$ and ${\bf Q_B}$ (e.g. two masses) by local interactions (by ${\bf Q_A}$ coupling to ${\bf M}$ and ${\bf M}$ to ${\bf Q_B}$, but without ${\bf Q_A}$ and ${\bf Q_B}$ interacting directly) then ${\bf M}$ must be non-classical. 

The theorem, in this general form, rests on two main principles, see \cite{MVPRD} for their details. One is the principle of locality, or {\sl no action at a distance}, that has to be satisfied by the dynamical theories describing both the mediator and the probes. (This principle then allows one to assume pairwise interactions of each quantum probe with the mediator, but not directly with each other).
 
The other is the principle of {\sl interoperability of information}, as defined in \cite{DEMA}. Informally, this principle expresses the requirement that the composite system of two systems that can each contain information, must be capable of containing information, too. Note that here `capable of containing information' has a precise operational definition, in terms of possible and impossible tasks. Specifically, given two systems that can, each, be prepared in any of a number of states that can be permuted in all possible ways, and copied, the interoperability principle requires that their composite system must, too, have a set of attributes that can all be permuted and copied in turn.  These two principles are obeyed by our best physical theories proposed so far, both classical and quantum: locality is necessary for the existence of independent subsystems, and both principles are necessary for testability.
By relying on these two principles, one need not assume a specific dynamics for the mediator in order to demonstrate the theorem about the witness. This is essential in order to interpret the test of non-classicality in gravity as ruling out, in case of observing entanglement, a large class of classical theories for the mediator, not just a specific decohered version of linear quantum gravity (as instead is assumed in \cite{SOUG, SOUG2}).  

In this sense, the theorem supporting these witnesses is like Bell's theorem: by violating Bell's inequalities, one can rule out a vast class of theories obeying general probabilistic assumptions; likewise, when observing entanglement between the masses, one can rule out all classical theories of gravity obeying the above-mentioned general principles: classical theories that are known (e.g., quantum field theory in curved spacetime, general relativity, some collapse theories, \cite{MVPRD}) and those that have yet to be discovered.

\section{Hall and Reginatto's proposed hybrid model}

There is a rich literature on hybrid quantum-classical systems, see e.g. \cite{HARE, BARCE, SAL, TER} and references therein.  Broadly speaking, they consist of inequivalent dynamical models where a classical and quantum sector coexist and can interact with each other, under given constraints that are required for consistency. 
Some of these models are not expressible as the fully decohered version of a quantum dynamical model; hence, it is of the essence to have our general theorem in order to rule them out as models that could still describe the generation of entanglement by a classical theory of gravity. Specifically, Hall and Reginatto \cite{HARE} asserted that there can be hybrid statistical quantum-classical models where the mediator {\bf M}, despite being classical, can still create entanglement between two quantum probes. Happily, as we shall explain, the proposed model in fact validates the witness. In short, this is because the model conceals a hidden non-locality in the configuration space dynamics, in contradiction with the locality assumption of our theorem; and when considering a local account, the model involves a non-classical mediator, with non-commuting variables that can be empirically accessed, as required by our witness.

Let us summarise here the model, following \cite{HARE, SAL}. Consider two quantum systems $Q$ and $Q'$ and a third one, $C$ which is classical. The statistical hybrid used by \cite{HARE}  is defined as follows: any point in configuration space is described by the triplet of real numbers $z=(q,q',x)$, representing each the position coordinate of  $Q$,  $Q'$ and $C$.  
The description based on the configurations space consists of the map of classical and quantum observables to their mean values
$$\begin{aligned}
f(x,\nabla_x S) &\to C_f[P,S] = \int dz P f(x,\nabla_x S) \\
\hat M &\to Q_{\hat M}[P,S] = \langle \psi |\hat M |\psi \rangle
\end{aligned}$$
which, for both sectors, they are functionals of $(P(z),S(z))$. We can introduce the hybrid Poisson brackets
$$ \{ A, B \}_H = \int dz P \Big [  \frac{\delta A}{\delta P} \frac{\delta B}{\delta S} - \frac{\delta A}{\delta S}\frac{\delta B}{\delta P} \Big ] $$
where $\frac{\delta A}{\delta f}$ denotes the variational derivative of the functional $A[f]$ with respect to $f$. Then, this induces an isomorphism from the algebra of quantum and classical sectors to the Poisson algebra of variables $C_f$ and $Q_{\hat M}$ in the configuration phase-space, defined as follows:
$$
\begin{aligned}
\{ C_f, C_g\}_H = C_{\{ f,g\}_P}\; , \; \{ Q_{\hat M}, Q_{\hat N}\}_H = Q_{[\hat M, \hat N]/i\hbar}
\end{aligned}
$$
with $f,g$ and $\hat M, \hat N$ being respectively some generic classical and quantum observables. Via this representation, the state of the composite system is specified by the probability density $P(z)$ and its conjugate density $S(z)$, while the time evolution driven by the Hamiltonian $H$ reads
\begin{equation}
\frac{\partial P}{\partial t} = \frac{\delta H}{\delta S}\;,\;\frac{\partial S}{\partial t} = -\frac{\delta H}{\delta P} \;.\nonumber
\end{equation}
Hall and Reginatto then analysed the interaction for the hybrid system $Q$, $Q'$ and $C$ given by:
\begin{equation}
H[P,S]= g_1 \int dq dq' dx P(\partial_q S) x +g_2 \int dq dq' dx P(\partial_x S) q'\; \label{EqHam}
\end{equation}
which represents, according to the authors, a {\sl pairwise} coupling between $Q$ and $C$ and then between $C$ and $Q'$, with $g_1$ and $g_2$ being the respective coupling constants. 
By applying the statistical hybrid model dynamical laws, one obtains that, for this particular Hamiltonian, the global state of the hybrid ensemble evolves as:
\begin{equation}
\psi_t (q,q',x) = \sqrt{P(q,q',x)}e^{iS(q,q',x)/\hbar}= e^{ -it H_{eff} }\psi_0(q,q',x), \label{EQ1}
\end{equation}
where $H_{eff}=(g_1\hat p\hat x+g_2 \hat q \hat k)$,  $\hat p=-i\hbar \partial_q$ and $\hat k={-i\hbar}\partial_x$ are the momenta operators of $Q$ and $C$ respectively. 
It is easy to check that by starting the three system in say a product state $\psi_0(q,q',x)= \psi_Q(q)\psi_{Q'}(q')\psi_C(x)$, it is possible to obtain a state $\psi_t (q,q',x) $ where $Q$ and $Q'$ are entangled, by evolving it as specified by $H$. 
How can this be, given the initial assumption about $C$ being classical? Hall and Reginatto conclude that this is a counterexample to the proposed witness. We shall now see that in fact this model validates it.

\section{What is really going on}

There are two explanations of what is going on in the example, each of which is compatible with the witness. The first explanation is manifest when considering the form of $H_{eff}$. 

Let us first clarify an important point. The state in \eqref{EQ1} is entangled if, when describing $Q$ and $Q'$ in the formalism of quantum theory, one can violate Bell's inequalities by measuring quantum observables of Q and Q'. So let us then assume this is the case, and consider carefully what this implies for the mediator $C$. As correctly pointed out by \cite{HARE}, $\hat k$ and $x$ are two variables of the sector $C$ that do not commute. Hence, the system $C$ is a non-classical system, in line with the witness' predictions.
However there is a subtlety: Hall and Reginatto claim that ${C}$'s non-classical variable $\hat k$ is not empirically accessible, in that it is an assumption of the statistical hybrid model that  ``classical observables" must take the form (for some well-behaved function $f$):

\begin{equation} \label{classical-obs}
C_f[P,S]=\int dx P f(x, \nabla_x S)\;.
\end{equation}
In other words, observables of ${\bf C}$ must be functions of $x$ and $\nabla_x S$ only. For this reason, according to Hall and Reginatto, $C$ is still a classical system, even when governed by a dynamical evolution that involves $\hat k$. But the ad-hoc requirement that observables of $C$ can assume only the form \eqref{classical-obs} is inconsistent with the very dynamical evolution proposed in the example; because that dynamical evolution is precisely a way to access $\hat k$ empirically. 

Specifically, upon confirming entanglement on $Q$ and $Q'$ by implementing a protocol to violate Bell's inequalities, one can empirically access the degree of freedom $\hat k$ of the system $C$, effectively performing its tomographic reconstruction. For, after the action of $H_{eff}$, the time-evolved quantum observables of Q and of Q'  become both functions of both variables $\hat{x}$ and $\hat{k}$. So, measuring the \textit{joint} observables of Q and Q' which confirm  entanglement, e.g. via an entanglement witness $\langle \mathcal{W} \rangle\approx \bra{\psi}q\hat p'+q'\hat p\ket{\psi}$, is an indirect measurement of the non-commuting degrees of freedom of the mediator ($x$ and $\hat k$), violating assumption \eqref{classical-obs} that $\hat k$ does not qualify as an 'observable' in the classical sense. Hence, the proposed hybrid model precisely adheres to the theorem in \cite{MAVE}: in order for $C$ to entangle $Q$ and $Q'$, it must be non-classical (i.e., its dynamics must allow for two non-commuting physical variables, $x$ and $\hat k$).  In passing, let us point out that a similar inconsistency is present in a different hybrid model, considered by \cite{SUD3}. There, a superselection rule is imposed on the classical system: a variable called $\hat \pi$ (which does not commute with the position variable of the classical system) is assumed not to be an observable; but at the same time it is allowed that the joint hamiltonian of the classical system and of the quantum system coupled to it depend on $ \hat \pi$. These two requirements are inconsistent, because it is possible to access $\hat \pi$ empirically via a protocol similar to that discussed in this note, where the classical system entangles two other quantum systems. 

More generally, what's observable is set by the allowed dynamical interactions in a given theory. If one assumes that Q and Q' become entangled, one also implicitly assumes that there are available interactions jointly on Q and Q' that can measure their joint observables to violate Bell inequalities. This is certainly the case at the level of the quantum observables and quantum commutators. This fact must be fully reproduced by the hybrid theory, if it is claimed to be able to bring about entanglement.

The other explanation for why the proposed counterexample is not a viable one, refers to the Hamiltonian on the configuration space, governing the hybrid ensemble, as in \eqref{EqHam}. That Hamiltonian is presented as providing a local coupling between $Q$ and $C$; and separately between $Q'$  and $C$. But, does it? As explained in \cite{HARE-Book}, it is possible to show that ensemble Hamiltonians of the form \eqref{EqHam} acting on a given subsystem have zero hybrid Poisson brackets with ensemble observables of other subsystems. But when the classical observables fall outside those particular cases, one has
    $$\{ Q_{\hat{M}} , C_f \}_H \neq 0$$ 
for $\hat{M}$ and $f$ observables in the quantum and classical sectors respectively, and for a generic wavefunction for Q, Q' and C. This precisely makes the theory disobey the principle of no-action at a distance, which is a requirement for the theorem in \cite{MVPRD}. The canonical commutator being zero for all variables of space-like separated system is necessary for a local theory of measurements and it is for guaranteed only in the condition of \textit{strong separability}, as already remarked \cite{SAL, SAVA, OLY}. But this condition is precisely violated in our experiment, given that Q and Q' must become entangled at the end of the interaction. 

One could attempt introducing a superselection rule that forces interactions to proceed always through configuration and momentum ensemble observables, as suggested in \cite{HARE-Book}. This move in itself is problematic, because this superselection rule does not have a corresponding rule in known quantum theory and standard classical mechanics, where all the interactions are allowed. Hence the correspondence between such superselected hybrid model, and existing well-corroborated physical theories would be violated.
 
Furthermore, even if one restrict attention to the superselected interactions (those which only go through configuration and linear momentum), the microscopic descriptors of the classical system $C$ constituting the hybrid ensembles already violate the axiom of no-action at a distance of the theorem in \cite{MAVE}. Hence the proposed hybrid model is ruled out a priori, according to this explanation. 

As in Bohmian mechanics, the problem is of course that the functional forms of $P(q,x,q')$ and $S(q,x,q')$ (when describing an entangled state) are responsible for the non-local behaviour; and the dynamical classical linear momenta are functions of $S$. The theorem in \cite{MAVE} requires no-action at a distance to hold for the complete descriptors of each subsystems: but this is violated by the fact that the momentum $\nabla_x S(q,x,q')$ of the classical sector, when Q and Q' are quantum entangled, can be modified by {\sl quantum operations} that involve only Q or only Q', both space-like separated from $C$.

This violation of no-action at a distance at the level of fundamental descriptors also disqualifies the hybrid model as a faithful description of quantum and classical systems constituting the ensembles. For the principle of no-action at a distance is satisfied by quantum theory (as it is manifest in the Heisenberg picture, \cite{DD}) and classical GR: so how can the hybrid ensemble approach (that purports to faithfully accommodate both) violate the principle? 

There is of course a third, trivial, explanation: that the state in equation \eqref{EQ1} isn't quantum-entangled after all, but it only formally resembles an entangled state. In this case, the entanglement cannot even be confirmed, i.e. the state does not violate Bell's inequalities -- hence it admits a local hidden variable model. In \cite{HARE}, it is never explained how to confirm entanglement in the hybrid model, modelling within the hybrid ensemble dynamics the measurements of {\sl local ensemble} observables of Q and Q' and classical communication only, as in standard protocols to violate Bell's inequalities. The problem of course is that a joint observable cannot be written as a product of {local} general \textit{ensemble observables} of Q and Q', such as $E_{\hat M}=\bra{\psi}\hat M\ket{\psi}$ or $E_{\hat M'}$, because in general (when Q and Q' are quantum-entangled)  $E_{\hat M}E_{\hat M'}\neq E_{\hat M\hat M'}$. According to this third explanation, no entanglement is formed between Q and Q', hence the theorem supporting the witness is trivially satisfied. 

Thus, to summarise: the proposed model is not a counterexample, but is in fact an example that adheres to the theorem supporting the witness.
In configuration space, the hybrid model violates locality; but locality is necessary to apply the witness, hence the model is not a counterexample.  
Indeed, according to the theorem supporting the non-classicality witness there may be non-local hybrid quantum-classical systems that can create entanglement, but such models are not physically plausible, given that locality is a pillar of both general relativity, non-relativistic quantum theory and quantum field theory. 

On the other hand, assuming that $Q$ and $Q'$ are entangled, if we look at the {\sl local} account of the proposed model which involves the Hamiltonian $H_{eff}$, we see that it includes two non-commuting variables for the mediator, $\hat k$ and $x$. Hence the mediator is non-classical and it provides an example of a non-classical system mediating entanglement. In addition, the variable $\hat k$ is, pace Hall and Reginatto, empirically accessible -- trough measurements that involve both $Q$ and $Q'$ and their mutual entanglement at the end of the entangling protocol. Thus, the hybrid model in this case validates the witness proposed in \cite{MAVE}, because it described a local generation of entanglement via a non-classical mediator.

We have therefore provided the promised vindication of the general witness of non-classicality and of the recently proposed schemes to test non-classical effects in gravity.

\maketitle
%\bibliography{biblio}% Produces the bibliography via BibTeX.

\begin{thebibliography}{31}%
\bibitem{MVPRD}  C. Marletto and V. Vedral, Phys. Rev. D{\bf },  2020.
\bibitem{SOUG} S. Bose, {\it et al.} Phys. Rev. Lett. {\bf 119}, 240401, 2017.
\bibitem{MAVE}  C. Marletto and V. Vedral, Phys. Rev. Lett. {\bf 119}, 240402, 2017.
\bibitem{HAL} M. Hall and M. Reginatto, Phys. Rev. A {\bf 78}, 042104, 2008.
\bibitem{HARE} M. Hall and M. Reginatto, J. Phys. A: Math. Theor. 51, 085303, 2018.
\bibitem{Collapse} G. Ghirardi and A. Bassi, The Stanford Encyclopedia of Philosophy (Summer 2020 Edition), Edward N. Zalta (ed.), 
\bibitem{DYS} T. Rothman and S. Boughn, Found Phys 36, 1801-1825 (2006). 
\bibitem{TER1} A. Ahmadzadegan, R. Mann, D. R. Terno, Phys. Rev A, {\bf 93}, 2016.
\bibitem{DEMA} D. Deutsch and C. Marletto, Proc. R. Soc. A {\bf 471}, 2174, 2014.
\bibitem{BARCE} C. Barcelo {\sl et al.}, Phys. Rev. A 86, 042120, 2012.
\bibitem{TER} A. Peres and D. Terno, Phys. Rev. A 63, 2001. 
\bibitem{SOUG2} R. Marshman, {\sl et al.} Phys. Rev. A 101, 052110, 2020.
\bibitem{HARE-Book} M. Hall and M. Reginatto, Ensembles on Configuration Space, Springer International Publishing (2016), Chapters 3-9.
\bibitem{SAL} L. L. Salcedo, Phys.Rev. A85, 022127, 2012.
\bibitem{SAVA} M. Hall, M. Reginatto and C. M. Savage, Physical Review A 86, 054101, 2012.
\bibitem{OLY} T. A. Olyinyk, Foundations of Physics volume 46, 2016.
\bibitem{SUD3} G. Sudarshan {\sl et al.}, Phys.Rev.D 20, 3081-3093, 1970.
\bibitem{DD} Deutsch, D., Hayden, P.  Proc. R. Soc. Lond. A456, 1759?1774, 2000.
\end{thebibliography}


{\bf Acknowledgements} \;\; The authors are grateful to David Deutsch and Vlatko Vedral for several comments; and to Michael Hall and Marcel Reginatto for useful criticism. This publication was made possible through the support of the Eutopia Foundation and of the ID 61466 grant from the John Templeton Foundation, as part of the The Quantum Information Structure of Spacetime (QISS) Project (qiss.fr). The opinions expressed in this publication are those of the authors and do not necessarily reflect the views of the John Templeton Foundation.

\end{document}